\documentclass[aps,pra,reprint,twocolumn,superscriptaddress]{revtex4-1}

\usepackage{amsmath,amssymb,amstext,amsthm}
\usepackage{graphicx}
\usepackage{epstopdf}
\usepackage{braket}
\usepackage{bbold}
\usepackage{bbm}
\usepackage{xcolor}
\definecolor{dred}{rgb}{.8,0.2,.2}
\definecolor{ddred}{rgb}{.8,0.5,.5}
\definecolor{dblue}{rgb}{.2,0.2,.8}

\begin{document}

%:Title Info
\title{Spin-Orbit States of Neutron Wavepackets}

% Joachim
\author{Joachim Nsofini} 
\affiliation{Department of Physics, University of Waterloo, Waterloo, ON, Canada, N2L3G1}
\affiliation{Institute for Quantum Computing, University of Waterloo,  Waterloo, ON, Canada, N2L3G1}
%Dusan
\author{Dusan Sarenac}
\affiliation{Department of Physics, University of Waterloo, Waterloo, ON, Canada, N2L3G1}
\affiliation{Institute for Quantum Computing, University of Waterloo,  Waterloo, ON, Canada, N2L3G1} 
%Chris
\author{Christopher J. Wood}
\affiliation{Department of Physics, University of Waterloo, Waterloo, ON, Canada, N2L3G1}
\affiliation{Institute for Quantum Computing, University of Waterloo,  Waterloo, ON, Canada, N2L3G1} 
\affiliation{IBM T.J. Watson Research Center, Yorktown Heights, NY 10598, USA}
% David
\author{David G. Cory}
\affiliation{Institute for Quantum Computing, University of Waterloo,  Waterloo, ON, Canada, N2L3G1} 
\affiliation{Department of Chemistry, University of Waterloo, Waterloo, ON, Canada, N2L3G1}
\affiliation{Perimeter Institute for Theoretical Physics, Waterloo, ON, Canada, N2L2Y5}
\affiliation{Canadian Institute for Advanced Research, Toronto, Ontario, Canada, M5G 1Z8}
% Arif
\author{Muhammad Arif}
\affiliation{National Institute of Standards and Technology, Gaithersburg, Maryland 20899, USA}
% Clark
\author{Charles W. Clark}
\affiliation{Joint Quantum Institute, National Institute of Standards and Technology and University of Maryland, Gaithersburg, Maryland 20899, USA}
% David
\author{Michael G. Huber}
\affiliation{National Institute of Standards and Technology, Gaithersburg, Maryland 20899, USA}
% Dmitry
\author{Dmitry A. Pushin}
\email{dmitry.pushin@uwaterloo.ca}
\affiliation{Department of Physics, University of Waterloo, Waterloo, ON, Canada, N2L3G1}
\affiliation{Institute for Quantum Computing, University of Waterloo,  Waterloo, ON, Canada, N2L3G1}

\begin{abstract}
We propose a method to prepare an entangled spin-orbit state between the spin and the orbital angular momenta of a neutron wavepacket. This spin-orbit state is created by passing neutrons through the center of a quadrupole magnetic field, which provides a coupling between the spin and orbital degrees of freedom. A Ramsey fringe type measurement is suggested as a means of verifying the spin-orbit correlations. 
\end{abstract}

\pacs{03.75.Be,42.50.Tx,03.65.Vf}

\maketitle

%==============================================================
% Introduction
%==============================================================
\section{Introduction }
\label{sec:intro}

Recently it was demonstrated that neutrons can support orbital angular momentum (OAM) states by using a spiral phase plate to write a helical wavefront onto a neutron beam \citep{Dima2015,Boyd2015}. This work is related to manipulating orbital angular momentum states of photons ~\cite{LesAllen1992,TwistedPhotons,PROP:PROP200410184,torner2005digital, Yao2011} and electrons \cite{OAMElexctrons,VGrillo2013,PhysRevLett.114.096102,Harris2015}. Neutrons have also an intrinsic spin of $\hbar/2$, and in this work we suggest a means of coupling the neutron spin and OAM to prepare an entangled spin-orbit state of a neutron wavepacket. This spin-orbit state could in principle be used for quantum metrology applications such as probing chiral and helical materials. 

For convenience let us consider a neutron beam propagating in the $z$-direction with momentum $k_z$, and the expectation values of momentum in the transverse plane equal to zero. The OAM operator in a cylindrical coordinate system $(r,\phi,z)$ is $L_z=-i \hbar \frac{\partial}{\partial \phi}$. The OAM eigenstates are a convenient basis for the neutron wavepacket when the coherence lengths in the transverse directions are equal $\sigma_x=\sigma_y\equiv \sigma_\bot$, where $\sigma_{x,y}=1/(2\Delta k_{x,y})$, and $\Delta k_{x,y}$ are the $x$ and $y$ spreads of the wavepacket's transverse momentum distributions.

Under this cylindrical symmetry the neutron wavefunction is separable in terms of spin and each of the cylindrical coordinates $\Psi_s(r,\phi,z)=R(r)\Phi(\phi)Z(z)\ket{s}$, where $s\in\{\uparrow={ 0 \choose 1}, \downarrow={ 1 \choose 0} \}$ specifies the neutron spin state along the quantization axis. With the standard deviation of momentum being constant in the transverse direction, the transverse wavefunction $R(r)\Phi(\phi)$ may be described in terms of solutions to the 2-D harmonic oscillator, and the longitudinal wavefunction $Z(z)$ treated as a Gaussian wavepacket. The eigenstates, denoted by $\ket{n_r,\ell,k_z,s}$, are specified by the radial quantum number $n_r$, the azimuthal quantum number $\ell$, the wave vector along the $z$ direction $k_z$, and the spin state $s$. 

The eigenstates in cylindrical coordinates are 
\begin{align}
\ket{n_r,\ell,k_z,s}=\mathcal{N} \xi^{|\ell|}e^{-\frac{\xi^2}{2}}\mathcal{L}_{n_r}^{|\ell|}\left(\xi^2\right)e^{i\ell\phi}Z(z)\ket{s},
\end{align}
where $\xi=r/\sigma_\perp$ is the rescaled radial coordinate, $\mathcal{N}=\frac{1}{\sigma_\perp}\sqrt{\frac{n_r !}{\pi(n_r+|\ell|)!}}$ is the normalization constant, $n_r\in(0,1,2...)$, $\ell\in (0, \pm 1, \pm2...) $, and $\mathcal{L}_{n_r}^{|\ell|}\left(\xi^2\right)$ are the associated Laguerre polynomials \cite{Olver:2010:NHMF}. The total neutron energy is
\begin{align}
E_T =\hbar\omega_\perp(2n_r+|\ell|+1)+\frac{\hbar^2k_z^2}{2m}-\vec{\mu}\cdot \vec{B},
\end{align}  
where $\vec{\mu}$ is the neutron magnetic dipole moment,  $\omega_{\perp}^2=\hbar/(2m\sigma_{\perp}^2)$, $m$ is the neutron mass, and $\vec{B}$ is the external magnetic field.

%-------------------------------------------------------
%: Neutron OAM States
%--------------------------------------------------------
\section{Neutron OAM States}

The first realization of a neutron OAM state was demonstrated using a spiral phase plate (SPP) \cite{Dima2015}. Before we considering spin-orbit states of neutrons it is useful to describe the action of a SPP in terms of orbital basis states. We may ignore the spin component here as the action of this spiral phase plate is spin independent. Consider a SPP of thickness $h(\phi)=h_0+h_s\phi/(2\pi)$, where $\phi$ is the azimuthal angle, $h_0$ is the base height, and $h_s$ is the step height. As a result of the optical potential \cite{Sears}, a neutron wavepacket propagating on axis through the SPP acquires a phase of $\alpha(\phi)=-Nb_c\lambda h(\phi)=\alpha_0+q\phi$, where $N b_c$ is the scattering length density of the SPP material, $\lambda$ is the neutron wavelength, $q=-Nb_c\lambda h_s/(2\pi)$ and the uniform phase $\alpha_0=-Nb_c\lambda h_0$.  The parameter $``q"$ is commonly referred to as the topological charge and it quantifies the nature of the singularity at the center \cite{Topphase}. Generally, when a plane wave propagates through such a topology, the wavefronts become $|q|$ intertwined helical surfaces with a helicity defined by the sign of $q$.

Let the incident neutron state carry well defined quantum numbers $n_{r_i}$ and $\ell_i$: $\ket{\psi_{in}}=\ket{{n_{r_i},\ell_i}}$, 
where we suppress the $k_z$ and $s$ labels as they are unaffected by the SPP.
To simplify we set $z=0$ at the exit of the SPP and we set $\sigma_\bot=1$. The state after the SPP can be expanded in terms of the basis functions
\begin{align}
\ket{\Psi_\mathrm{SPP}}&
=e^{iq\phi}\ket{\psi_{in}}
=\sum_{n_r=0}^{\infty}\sum_{\ell=-\infty}^{\infty}  C_{n_r\ell} \ket{n_r,\ell},
\end{align}
with the coefficients
\begin{align}
C_{n_r,\ell}&= \int_0^\infty \hspace{-0.75em}dr\int_0^{2\pi}\hspace{-0.75em}d\phi  \ r\langle n_r,\ell\ket{\Psi_\mathrm{SPP}} .
\end{align}

%
%
%%%%%================================================================================
\begin{figure}
\center
\includegraphics[width=\columnwidth]{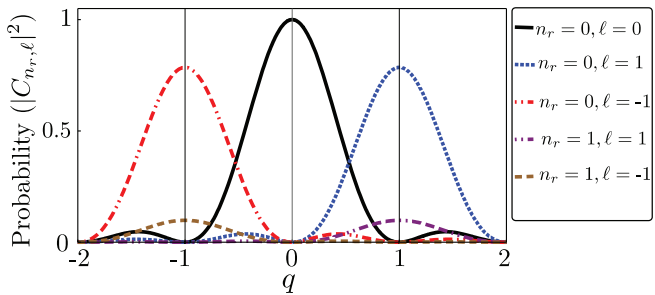}
\caption{The probabilities for each of the $\ell=0,1,-1$ and $n_r=0,1$ states when a neutron wavepacket with no OAM ($n_{r_i}=\ell_i=0$) passes through a spiral phase plate with a topological charge $q$.}
 \label{Fig:CoeffSPP}
\end{figure}
%%%================================================================================
%
When the incoming neutrons have zero OAM ($n_{r_i}=\ell_i=0$), the coefficients are
\begin{align}
C_{n_r,\ell}
&= \begin{cases}
e^{iq\pi}\text{sinc}(q\pi) \qquad \text{for }n_r=\ell=0 \\
\frac{ \frac{|\ell|}{2}\Gamma\left(1+\frac{|\ell|}{2}\right)}{\sqrt{n_r!(n_r+|\ell|)!}} e^{i(q-\ell)\pi}\text{sinc}[(q-\ell)\pi] & \text{otherwise}
\end{cases}
\label{Eqn:Coef00}
\end{align}
where $\Gamma\left(1+|\ell|/2\right)$ is the gamma function \cite{Olver:2010:NHMF}. When the incoming state has a definite orbital quantum number $\ell_i$, the output state is a state with definite orbital quantum number $\ell_i+q$.  

Fig.~\ref{Fig:CoeffSPP} shows the probabilities $(|C_{n_r,\ell}|^2)$ for $n_r=0,1$ and $\ell=0,1,-1$. From Eq.~(\ref{Eqn:Coef00}) we see that $C_{n_r,\ell=0}\neq 0$ only when $n_r=0$. From Fig.~\ref{Fig:CoeffSPP} we see that when a neutron wavepacket with zero OAM passes through a SPP, the OAM quantum number of the neutron wavepacket is incremented by the topological charge ($q$) of the SPP. The radial quantum number of the outgoing wavepacket can take any allowed value, the most probable one is $n_r=0$ for small $q$-values.  
If we consider, for example, a topological charge of $q=+1$ then the state after the SPP is 
\begin{align}
\ket{\Psi_\mathrm{SPP}}
=\sum_{n_r=0}^{\infty} &\sqrt{\frac{\pi}{16n_r!(n_r+1)!}}  \ket{n_r,1}.
\end{align}
Hence a SPP provides control over the orbital quantum number. 
$^3$He neutron counting detectors do not distinguish different radial states and so the effect of measurement traces over the radial quantum number.

Below we propose and analyze a method to create a neutron spin-orbit state over the coherence length of a  neutron wavepacket. The spin independent optical phase from the SPP is replaced by a spatially dependent spin rotation. The OAMs are generated as a result of the topological phase arising from the spin rotations induced by a quadrupole magnetic field. The resulting state is a spin-orbit state.

%----------------------------------------------------
%                 Neutron Spin-orbit states
%----------------------------------------------------
%
\section{Neutron Spin-orbit states}

Consider a neutron wavepacket spin polarized along the $z$-direction, traveling through a quadrupole magnetic field geometry ${\partial B_x}/{\partial y}=-{\partial B_y}/{\partial x}$. The magnetic field vector is given by $\vec{B}=-|\nabla B|r(\cos(q\phi),\sin(q\phi),0)$ where $|\nabla B|$ is the quadrupole gradient. The topological charge of the quadrupole is $q=-1$. Note that the magnitude of the magnetic field varies radially, while the direction changes azimuthally. The Hamiltonian of the neutron inside the quadrupole field can be parametrized by $H=-\vec{\hat{\sigma}}\cdot \vec{B}\gamma\hbar/2$ where $\vec{\hat{\sigma}}$ corresponds to the Pauli matrices and $\gamma=-2|\vec{\mu}|/\hbar$ is the neutron gyromagnetic ratio. The effect of this Hamiltonian in generating OAM is similar to those used in optics to generate OAM based on Pancharatnam-Berry geometrical phases \cite{Liberman1992,Bomzon2002,Lmarrucci2006,Mair2001, EkarimiSpinorbit1, SpinOrbitatoms, cardano2015spin}, and as shown recently for generating OAM with electrons through a type of Wien filter \cite{Ekarimi2012,VGrillo2013,Harris2015}.

Assuming the neutron is traveling along the quadrupole axis, the time the neutron spends inside the quadrupole magnetic field is $t_Q=l_Q/v_z$, where $l_Q$ is the length of the quadrupole and $v_z=2\pi\hbar /(m\lambda)$ is the neutron velocity. Ignoring the small radial neutron path displacement due to the gradient, the operator on the spin is
\begin{align}
U_Q
&=\cos\left(\frac{\pi r}{2r_c}\right) \mathbb{1}+i\vec{n}\cdot\vec{\hat{\sigma}}
\sin\left(\frac{\pi r}{2r_c}\right),
\label{Eqn:QuadDef}
\end{align}
where $\vec{n}\cdot\vec{\hat{\sigma}}=\left(\hat{\sigma}_x\cos\phi
-\hat{\sigma}_y\sin\phi\right)$, and we have re-parametrized the operator using the radius $r_c$ at which the spin undergoes a spin flip after passing through the quadrupole ${\gamma|\nabla B|r_c l_Q/v_z}=\pi$. 
The action of the quadrupole depends on its length, the gradient strength, and the neutron wavelength.

Defining raising and lowering OAM operators $l_\pm =e^{\pm i\phi}$ and spin operators $\hat{\sigma}_\pm =(\hat{\sigma}_x \pm i\hat{\sigma}_y)/2$, the operator of the quadrupole in Eq.~(\ref{Eqn:QuadDef})  becomes
\begin{align}
U_Q&=\cos\left(\frac{\pi r}{2r_c}\right) \mathbb{1}+
i\sin\left(\frac{\pi r}{2r_c}\right)(l_+\hat{\sigma}_++l_-\hat{\sigma}_-).
\label{Eq:QuadOAMOP}
\end{align}
The second term of Eq.~(\ref{Eq:QuadOAMOP}) is an entangling operation between spin and orbital momenta.
Hence passage of a neutron wavepacket through a quadrupole has the potential to entangle the spin and orbital degrees of freedom though we must also consider changes to the radial quantum number.

Consider the case where a spin-up polarized neutron is initially in a well defined OAM eigenstate $\ket{\psi_{in}}=\ket{{n_{r_i},\ell_i},\uparrow}$ and is passing on axis through the quadrupole. If we ignore the small change in $k_z$, $\Delta k_z$, and $\sigma_\perp$ as the wavepacket propagates through the quadrupole, the state after the quadrupole can be expanded in the basis functions as
\begin{align}
\ket{\Psi_{Q}}
&=\sum_{n_r=0}^{\infty}\sum_{\ell=-\infty}^{\infty}\left( C_{n_r,\ell,\uparrow}\ket{n_r,\ell,\uparrow}+iC_{n_r,\ell,\downarrow}\ket{n_r,\ell,\downarrow }\right).
\label{Eqn:QuadState}
\end{align}
The coefficients in Eq.~\eqref{Eqn:QuadState} are given by
\begin{align}
C_{n_r,\ell,\uparrow}
&=\int_0^\infty \hspace{-0.75em} d\xi \int_0^{2\pi}\hspace{-0.75em} d\phi\ 
	\langle n_r, \ell \ket{n_{r_i},\ell_i} \xi \cos\left(\frac{\pi \sigma_\perp}{2r_c}\xi\right)
	\label{Eqn:Quadcoeffup}\\
C_{n_r,\ell,\downarrow}
&= \int_0^\infty \hspace{-0.75em} d\xi \int_0^{2\pi}\hspace{-0.75em} d\phi\ 
	\langle n_r, \ell \ket{n_{r_i},\ell_i} \xi e^{i\phi}\sin\left(\frac{\pi \sigma_\perp}{2r_c}\xi\right)
\label{Eqn:Quadcoeffdown}
\end{align}
Integrating over $\phi$ selects $\ell=\ell_i$ for the spin-up coefficients, and $\ell=\ell_i+1$ for the spin-down coefficients. This simplifies Eq.~\eqref{Eqn:QuadState} to
\begin{align}
\ket{\Psi_{Q}}
&=\sum_{n_r=0}^{\infty}\left( C_{n_r,\ell_i,\uparrow}\ket{n_r,\ell_i,\uparrow}+
iC_{n_r,\ell_i+1,\downarrow}\ket{n_r,\ell_i+1,\downarrow }\right).
\end{align}
Note that this coupling between spin and OAM can easily be seen from the quadrupole operator in Eq.~(\ref{Eq:QuadOAMOP}). 

\begin{figure}
\center
\includegraphics[width=\columnwidth]{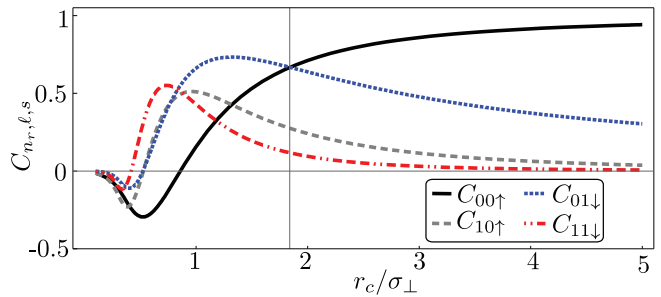}
\caption{The coefficients $C_{n_r,\ell,s}$ of the spin-orbit state for $n_r=0$ and $n_r=1$ as a function of $r_c/\sigma_\bot$. The input state is $n_{r_i}=\ell_i=0$. The vertical line at $r_c/\sigma_\bot=1.82$ corresponds to the point of maximum concurrence for the $n_r=0$ subspace (see Fig.~\ref{Fig:NormCon}). Strong quadrupole fields correspond to $r_c\rightarrow 0$ while no quadrupole $r_c \rightarrow \infty$.}
 \label{fig:Coeffs}
\end{figure}

The coefficients $C_{n_r,\ell_i,\uparrow}$ and $C_{n_r,\ell_i+1,\downarrow}$ are real for all values of $r_c/\sigma_\perp$.  The various coefficients  $C_{n_r,\ell,s}$ are plotted in Fig.~\ref{fig:Coeffs}, given an input state with $n_{r_i}=\ell_i=0$. The ratio $r_c/\sigma_\perp$ quantifies the action of the quadrupole on the neutron wavepacket. The strong quadrupole fields regime correspond to $r_c\rightarrow 0$ and the weak quadrupole regime to $r_c \rightarrow \infty$. It can be verified that
$
\sum_{n_r=0}^{\infty}(C_{n_r,\ell_i,\uparrow}^2
+C_{n_r,\ell_i+1,\downarrow}^2)=1.
$
%
%
%
%%%%%%================================================================================
%%%%%%================================================================================

%%%%%-----------------------------------------------------------
%%%Characterizing the Spin-Orbit States
%%%%%----------------------------------------------------------
\subsection{Characterizing the Spin-Orbit States}
Neutron interferometers have been used to demonstrate single-particle entanglement between different degrees of freedom, such as spin + path and spin + energy, and have supported extensive studies of quantum contextuality \cite{Hasegawa2003, Bartosik2009,Erdosi2013,Wood_2014,Klepp_2014}.
A useful measure of entanglement for a bipartite quantum system is the \emph{concurrence}  \cite{Wooters1998,Rungta2001,Fei2001}, which is equal to 1 when the entanglement is maximum and 0 when the state is separable. For a bipartite mixed state $\rho_{SO}$, the concurrence is given by
\begin{align}
{\mathcal{C}(\rho_{SO})}=max\{0,\lambda_1-\lambda_2-\lambda_3-\lambda_4\},
\label{Eqn:ConMix}
\end{align}
where the $\lambda_i$'s are the eigenvalues, sorted in descending order, of  $\sqrt{\sqrt{\rho_{SO}}(\sigma_y\otimes\sigma_y) \rho_{SO}^*(\sigma_y\otimes\sigma_y)\sqrt{\rho_{SO}}}$, and $\rho_{SO}^*$ is the complex conjugate of $\rho_{SO}$.
For a pure state $\rho_{SO}=\ket{\psi_{SO}}\!\!\bra{\psi_{SO}}$, Eq.~\eqref{Eqn:ConMix} reduces to
\begin{align}
\mathcal{C} (\ket{\psi_{SO}})&=\sqrt{2\left(1-\text{Tr}[\rho_{S}^2]\right)},
\label{Eqn:ConC}
\end{align}
where $\rho_{S}=\text{Tr}_{O}[\ket{\psi_{SO}}\!\!\bra{\psi_{SO}}]$ is the reduced density matrix obtained by tracing over the subsystem $S$ (or equivalently tracing could be over subsystem $O$).

\begin{figure*}
\center
\includegraphics[width=17.2cm]{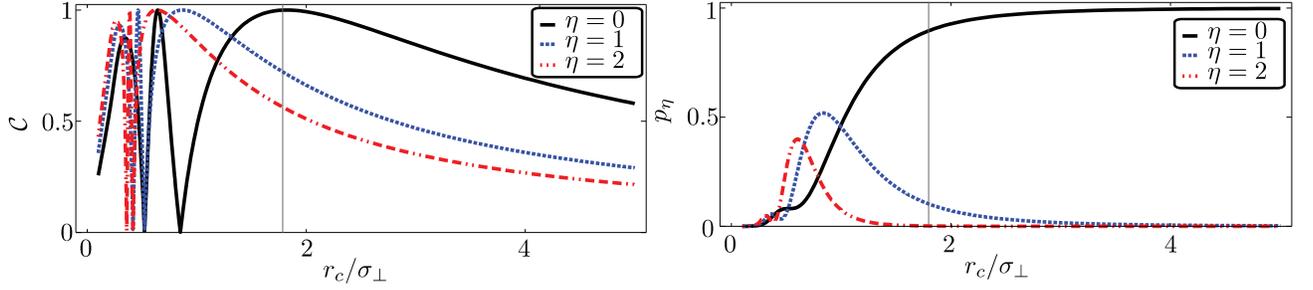}
\caption{Concurrence (on the left) of the spin-orbit state for the filtered $\eta=0,1,2$ subspaces, and the probability (on the right) of the given  $\eta=0,1,2$ subspaces. The vertical line at $r_c/\sigma_\perp=1.82$ corresponds to the point of maximum concurrence for the $\eta=0$ subspace.}
%\twocolumngrid
 \label{Fig:NormCon}
\end{figure*}

Let us first consider the entanglement of the spin-orbit neutron state in the case where we filter on a single radial quantum number $n_r=\eta$. In this case the renormalized spin-orbit state is a pure state
\begin{align}
\ket{\psi_\eta}&=\frac{1}{\sqrt{p_{\eta}}}\left(C_{\eta,\ell_i,\uparrow}\ket{\ell_i,\uparrow}+iC_{\eta,\ell_i+1,\downarrow} \ket{\ell_i+1,\downarrow}\right),
\label{Eqn:SpinObitNorm}
\end{align}
where $p_\eta$ is the probability of the the wave-packet being in the specific $n_r=\eta$ subspace:
\begin{align}
p_{\eta}&=C_{\eta,\ell_i,\uparrow}^2
+C_{\eta,\ell_i+1,\downarrow}^2.
\label{Eqn:probnr}
\end{align}
The concurrence of the $\ket{\psi_\eta}$ and probability coefficients $p_\eta$ as a function of $r_c/\sigma_\perp$  are shown in Fig.~\ref{Fig:NormCon} for the radial subspaces $\eta=0,1,2$. The concurrence of the spin-orbit state obtained by passing through a quadrupole is maximized for the $\eta=0$ radial subspace when there is a spin flip at $\sim 1.82$ times the coherence length of the wavepacket. This condition is represented by the vertical line in Fig.~\ref{fig:Coeffs} and Fig.~\ref{Fig:NormCon}.

%%%%%%================================================================================
%%%================================================================================
%

\begin{figure}
\center
\includegraphics[width=\columnwidth]{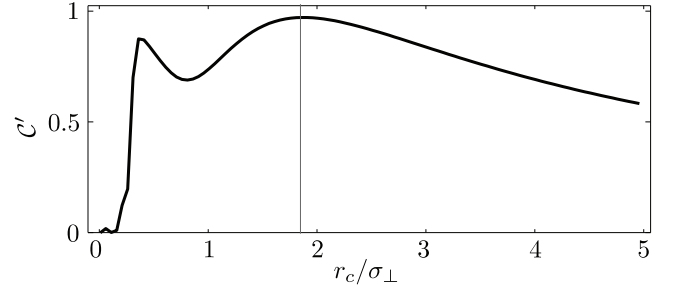}
\caption{Concurrence of the spin-orbit state obtained by tracing the radial subspace. The vertical line at $r_c/\sigma_\perp=1.82$ corresponds to the point of maximum mixed state concurrence of 0.97. The concurrence does not go to 1 because the traced state is not pure.}
 \label{Fig:TrCon}
\end{figure}

Next we consider the case where the neutron capture cross-section of the detector is independent of the $n_r$ subspace. For $n_{r_i}=\ell_i=0$, the spin-orbit density matrix obtained by tracing over the radial degree of freedom is
\begin{align}
\rho_{SO}&=\sum_{n_r=0}^{\infty}\left[C_{n_r,0,\uparrow}^2\ket{0,\uparrow } \bra{0,\uparrow } 
+iC_{n_r,0,\uparrow} C_{n_r,1,\downarrow} \ket{0,\uparrow } \bra{1\downarrow }\right. \nonumber\\
 &\left.\qquad -i C_{n_r,0,\uparrow}C_{n_r,1,\downarrow} \ket{1,\downarrow }\bra{0,\uparrow }
+C_{n_r,1,\downarrow}^2 \ket{1,\downarrow }\bra{1,\downarrow }\right ].
\label{Eqn:SpinOrbitTracenr}
\end{align}
This reduced state is not a pure state ($Tr[\rho_{SO}^2]\neq 1$). The concurrence of this mixed spin-orbit state can be computed using Eq.~\eqref{Eqn:ConMix} and the resulting value is shown in Fig.~\ref{Fig:TrCon}. We find that the maximum value of concurrence is $\mathcal{C}(\rho_{SO})= 0.97$ and it occurs at $r_c/\sigma_\bot=1.82$. Hence even after averaging over all radial subspaces the spin-orbit state is still highly entangled.

\section{Proposed Experimental Implementation}

To experimentally implement this proposal a quadrupole magnet can be constructed from correctly orientated discrete NdFeB magnets. A 10 cm long quadrupole with a gradient of 13.8 T/cm  would be required to satisfy the $r_c=1.82\sigma_\bot$ condition for neutrons with a typical transverse coherence length of $\sigma_\perp=100$ nm \citep{Sam,pushin_prl_coherence:250404} and a wavelength of 0.271~nm. With the 0.7 T surface field of  NdFeB magnets this gradient corresponds to an inner quadrupole gap of around 1 mm. In such an experimental setup, the concurrence (Eq.~(\ref{Eqn:ConC})) of the $\eta=0,1,2$ filtered states is 1, 0.77, and 0.55 respectively, and the traced concurrence (Eq.~(\ref{Eqn:ConMix})) is 0.97.

\begin{figure}
\center
\includegraphics[width=\columnwidth]{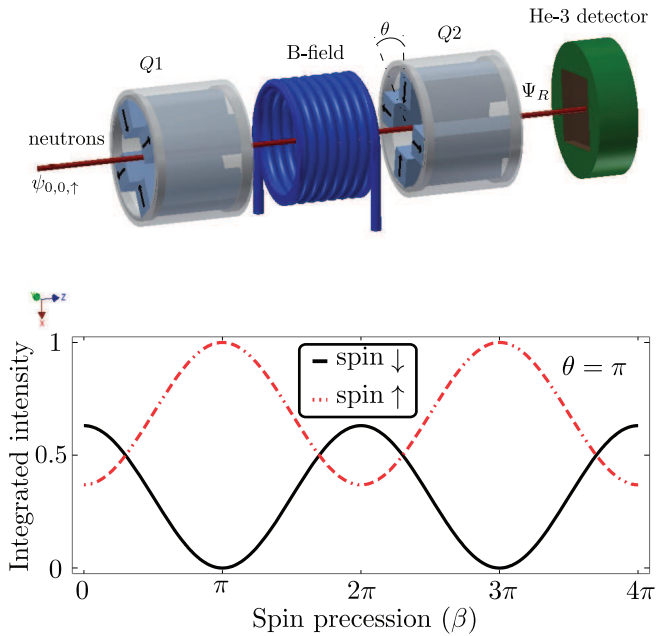}
\caption{The top figure is the setup for the spin-orbit Ramsey fringe experiment. The arrows on the magnets depict the quadrupole geometry. The bottom figure is the integrated intensity at the output for the spin-up and spin-down neutrons as a function of the spin precession ($\beta$) inside the solenoid. The rotation of the second quadrupole is set to $\theta=\pi$.   An identical plot can be obtained when $\beta=\pi$ and the quadrupole rotation is varied. This variation of the intensity is an indication of the correlations between the spin and OAM. The phase induced by the spin rotation can be compensated by the rotation of the quadrupole.}
 \label{Fig:Ramsey}
\end{figure}

The successful preparation of the entangled state could be verified by using a Ramsey Fringe experiment~\cite{Ramsey}. For the experiment we require a polarized neutron beam, two quadrupole magnets and a solenoid between them (see Fig.~\ref{Fig:Ramsey}). The first quadrupole creates the spin-orbit state. The solenoid provides a uniform magnetic field along the spin quantization axis and introduces a phase shift, $\beta$, in the spin degree of freedom. The corresponding operator is $
U_z(\beta)=\cos\left({\beta}/{2}\right) \mathbb{1}+
i\sin\left({\beta}/{2}\right)\hat{\sigma}_z.
$
The second quadrupole can be rotated by angle $\theta$ and can act as an inverse operator of the first quadrupole:
\begin{align}
U_{Q2}(\theta)&=\cos\left(\frac{\pi r}{2r_c}\right) \mathbb{1}+
i\sin\left(\frac{\pi r}{2r_c}\right)\left[e^{-i\theta}l_+\hat{\sigma}_++e^{i\theta}l_-\hat{\sigma}_-\right]
%\label{Eqn:UQ2}
\end{align}
%
%%%%%%================================================================================
%%%================================================================================
%
%
With the setup shown in Fig.~\ref{Fig:Ramsey}, when the input state is  $\ket{0,0,\uparrow}$, the state at the exit (global phase excluded) is
\begin{align}\nonumber
\ket{\Psi_{R}}&=U_{Q2}(\theta) U_z(\beta) U_Q\ket{0,0,\uparrow}\\\nonumber
&=\left[\cos\left(\frac{\pi r}{r_c}\right)\cos\left(\frac{\beta-\theta}{2}\right)-
i\sin\left(\frac{\beta-\theta}{2}\right)\right]\ket{0,0,\uparrow}
\\ & \qquad -i \sin\left(\frac{\pi r}{r_c}\right)\cos\left(\frac{\beta-\theta}{2}\right)e^{i\phi}\ket{0,0,\downarrow }
%\label{eqn:Ramsey}.
\end{align}
The integrated intensities at the output are
\begin{align}\nonumber
\overline{I_{\uparrow}}(\beta,\theta) &=
	\int_0^\infty \hspace{-0.75em}dr \int_0^{2\pi} \hspace{-0.7em}d\phi \ 
	r\,|\langle \uparrow\ket{\Psi_{R}}|^2 \\
&=1-\frac{\pi\sigma_\perp}{r_c} F\left(\frac{\pi\sigma_\perp}{r_c}\right)\cos^2\left(\frac{\beta-\theta}{2}\right)\label{Eqn:intenRamseyDn}\\\nonumber
\overline{I_{\downarrow}}(\beta,\theta)
&=\int_0^\infty \hspace{-0.75em}dr \int_0^{2\pi} \hspace{-0.75em}d\phi \ 
	r |\langle\downarrow\ket{\Psi_{R}}|^2 \\
&=\frac{\pi\sigma_\perp}{r_c}F\left(\frac{\pi\sigma_\perp}{r_c}\right)\cos^2\left(\frac{\beta-\theta}{2}\right),
\label{Eqn:intenRamseyUp}
\end{align}
where $F(\pi\sigma_\perp/r_c)$ is Dawson's intergral \cite{Olver:2010:NHMF}. 
The integrated spin intensities at the output (Eq.~(\ref{Eqn:intenRamseyDn}) \& (\ref{Eqn:intenRamseyUp})) show the same behaviour if $\beta$ is varied for fixed $\theta$, and if $\theta$ is varied for fixed $\beta$. The fact that the phase induced by the spin rotation can be compensated by the rotation of the orbital state is an indication of the spin-orbit entanglement. The bottom part of Fig.~\ref{Fig:Ramsey} displays the spin-dependent integrated intensity for $\beta$ varied with $\theta=\pi$ and with $r_c/\sigma_\perp=1.82$. Note that the amplitude of the oscillations of the integrated intensity is not 1 because the spin-orbit state obtained by tracing the radial degree of freedom is not pure.
%
%%%%%================================================================================
%%%================================================================================

\section{Conclusion}

We have proposed a method for preparing spin-orbit states of neutron wavepackets using a quadrupole magnetic field.
We have also demonstrated that the spin-orbit state would be entangled, and that this entanglement is maximized for certain values of the coherence length and quadrupole magnetic field strength. Successful realization of the spin-orbit states  will provide an opportunity to use neutrons as a probe of chiral and helical materials. For example, these unique spin-orbit coupled states may be used to study chiral magnetic materials and skyrmions.

\section{Acknowledgements}
This work was supported by the Canadian Excellence Research Chairs (CERC) program, the Natural Sciences and Engineering Research Council of Canada (NSERC) Discovery program, Collaborative Research and Training Experience (CREATE) program, and the National Institute of Standards and Technology (NIST) Quantum Information Program. The authors are greatful to S. A. Werner for useful discussions.

%==============================================================
%:References
%==============================================================
%\bibliographystyle{unsrt}
\bibliographystyle{apsrev4-1}

\bibliography{OAM.bib}

\end{document}